\documentclass[sigconf]{acmart}
\AtBeginDocument{%
  }

\usepackage{subcaption}
\usepackage{listings}
\usepackage{xcolor}
\usepackage{booktabs}

\lstdefinestyle{promptstyle}{
  basicstyle=\ttfamily\small,
  columns=fullflexible,
  breaklines=true,
  frame=single,
  rulecolor=\color{black!20},
  showstringspaces=false,
  xleftmargin=0.5em,
  xrightmargin=0.5em,
  aboveskip=0.75em,
  belowskip=0.75em
}

\setcopyright{acmlicensed}
\copyrightyear{2018}
\acmYear{2018}
\acmDOI{XXXXXXX.XXXXXXX}
\acmConference[Conference acronym 'XX]{Make sure to enter the correct
  conference title from your rights confirmation email}{June 03--05,
  2018}{Woodstock, NY}
\acmISBN{978-1-4503-XXXX-X/2018/06}

\begin{document}

\title{Agora: Teaching the Skill of Consensus-Finding with AI Personas Grounded in Human Voice}

\author{Prerna Ravi}
\affiliation{%
  \institution{Massachusetts Institute of Technology}
  \city{Cambridge}
  \state{MA}
  \country{USA}
}
\email{prernar@mit.edu}

\author{Om Gokhale}
\affiliation{%
  \institution{Massachusetts Institute of Technology}
  \city{Cambridge}
  \state{MA}
  \country{USA}
}
\email{ogo@mit.edu}

\author{Suyash Pradeep Fulay}
\affiliation{%
  \institution{Massachusetts Institute of Technology}
  \city{Cambridge}
  \state{MA}
  \country{USA}
}
\email{sfulay@mit.edu}

\author{Eugene Yi}
\affiliation{%
  \institution{University of Oxford}
  \city{Oxford}
  \country{United Kingdom}
}
\email{eyi@mit.edu}

\author{Deb Roy}
\affiliation{%
  \institution{Massachusetts Institute of Technology}
  \city{Cambridge}
  \state{MA}
  \country{USA}
}
\email{dkroy@mit.edu}

\author{Michiel Bakker}
\affiliation{%
  \institution{Massachusetts Institute of Technology}
  \city{Cambridge}
  \state{MA}
  \country{USA}
}
\email{bakker@mit.edu}

\renewcommand{\shortauthors}{Fulay et al.}

\begin{abstract}
Deliberative democratic theory suggests that civic competence—the capacity to navigate disagreement, weigh competing values, and arrive at collective decisions—is not innate but developed through practice. Yet opportunities to cultivate these skills remain limited, as traditional deliberative processes like citizens' assemblies reach only a small fraction of the population. We present Agora, an AI-powered platform that uses LLMs to organize authentic human voices on policy issues, helping users build consensus-finding skills by proposing and revising policy recommendations, hearing supporting and opposing perspectives, and receiving feedback on how policy changes affect predicted support. In a preliminary study with 44 university students, access to the full interface with voice explanations—as opposed to aggregate support distributions alone—significantly improved self-reported perspective-taking and the extent to which statements acknowledged multiple viewpoints. These findings point toward a promising direction for scaling civic education.

\end{abstract}



\maketitle

\section{Introduction}


One premise of deliberative democratic theory is that civic competence is not innate, but developed through practice \cite{McDevitt01072006}.
The capacities that enable citizens to navigate disagreement, weigh competing values, and arrive at collective decisions—what \citet{kirlin2003role} terms "civic skills"—are competencies that allow individuals to become participants in democratic processes rather than observers. These skills are not fixed traits; they must be learned and practiced \cite{kirlin2003role}.


Framing deliberation as a skill rather than a disposition has important implications: if democratic capacities are learnable, the question becomes how to foster their development. Civic education highlights several relevant competencies, including communication, collective decision-making, and critical thinking \cite{kirlin2003role}. Yet opportunities to practice these capacities remain limited in everyday civic life. While well-designed deliberative processes—citizens' assemblies, deliberative polls, and forums—can improve the quality and consistency of participants' opinions \cite{fishkin2009people}, they reach only a small fraction of the public. In response to this challenge, we present Agora, a platform that turns passive exposure to an opinion landscape into active practice. Users propose a policy recommendation, hear authentic perspectives, and receive feedback as they iteratively revise their position. This cycle mirrors core operations of deliberative skill development, including weighing trade-offs, engaging disagreement, and refining proposals in light of others' reasoning \cite{shaffer_delib, Maia_2024}.

Our approach builds on two ideas. First, drawing on Dewey's pragmatist education philosophy, we argue democratic competencies develop through experience and feedback—not instruction alone \cite{dewey1916democracy}. Second, we propose deliberative skills can be developed through mediated exposure to authentic diverse perspectives \cite{considerit, kim2021starrythoughts}, suggesting that LLM-powered tools that organize and scaffold real voices could cultivate these skills at scale.

Existing digital deliberation systems have helped people explore opinion landscapes, reflect on trade-offs, and participate in structured discussion, while recent AI systems can help mediate disagreement or synthesize common ground. However, these systems are typically designed either to support a deliberative process or to generate a collective output, rather than to help individuals \emph{practice} the skill of crafting broadly acceptable policies while engaging authentic human perspectives. We focus on this missing educational layer: whether AI can scaffold consensus-finding as a learnable civic skill.

We evaluate this approach through a study in which participants use the dynamic Agora interface to try to find consensus on two policy issues in the United States: the minimum wage and domestic vs. foreign hiring priorities. We assess the tool's effects on fostering internal deliberation, perspective-taking, and learning, and improving the quality of generated consensus policies. In a randomized experiment with 44 students, the full version of the tool significantly improved self-reported perspective-taking, as well as the quality of written consensus statements along the dimension of perspective acknowledgment.

Our contribution is threefold. First, we introduce Agora, a system for practicing consensus-finding through iterative policy drafting, exposure to authentic perspectives, and immediate feedback on predicted support. Second, we show how LLMs can be used not to replace public voice, but to organize human interviews into a deliberative learning environment. Third, through a preliminary randomized study, we provide evidence that access to voice-grounded explanations can improve self-reported deliberative learning outcomes and the extent to which consensus statements themselves acknowledge different perspectives.

\section{Related Work}

\subsection{Deliberative Democracy and Pedagogy}

A core premise of this paper is that deliberative capacity is not a fixed disposition but something that can be developed through structured practice. Work in civic education and deliberative pedagogy argues that democratic competence emerges through repeated opportunities to weigh trade-offs, engage disagreement, justify one's views, and revise positions in light of competing perspectives \cite{McDevitt01072006,Maia_2024,shaffer_delib}. This perspective is important for our work because it shifts the goal of civic technologies away from merely informing users and toward helping them practice the underlying skills of collective judgment.

At the same time, the deliberative democracy literature has shown that high-quality deliberation is difficult to scale institutionally. Mini-publics, such as citizens' assemblies and deliberative polls, can improve the quality of public reasoning, but they typically involve limited numbers of participants and substantial organizational overhead \cite{fishkin2009people,setala2018mini,CitizensAssembliesandDemocracy}. Fishkin characterizes this as a trilemma between political equality, mass participation, and deliberation quality: democratic processes can usually realize only two of these ideals at once \cite{fishkin_trilemma}. Agora is motivated in part by this challenge. Rather than replacing citizens' assemblies or formal public consultation, it explores whether AI-mediated environments can scale \emph{practice} in consensus-finding by giving users repeated exposure to disagreement, feedback, and revision.

This framing also helps clarify the contribution of our system. We do not claim that brief interactions with an interface are equivalent to full interpersonal deliberation. Instead, we position Agora as a preparatory or complementary form of deliberative learning: an environment where users can iteratively propose, test, and revise policy ideas while engaging with authentic perspectives from others \cite{McDevitt01072006,Maia_2024}. In this sense, our work extends deliberative pedagogy into an AI-mediated setting.

\subsection{AI-Mediated Deliberation and Digital Representation}

A growing body of work examines how AI systems might support collective decision-making, not only by summarizing discourse, but also by mediating disagreement, eliciting preferences, and representing publics at scale. Tessler et al.\ show that AI-mediated deliberation can help groups identify common ground, with participants often preferring AI-generated consensus statements to those produced by human mediators on dimensions such as clarity, informativeness, and perceived impartiality \cite{habermas}. Related work on Polis argues that LLMs may help scale deliberation by clustering, summarizing, and surfacing patterns in large volumes of public input, while also raising concerns about bias, misrepresentation, and excessive concentration of interpretive power \cite{small2023opportunities}.

Recent work has also started to map the broader design space of AI in democratic processes. McKinney identifies eleven possible applications of AI across the lifecycle of citizens' assemblies, from agenda-setting support to generating consensus statements \cite{mckinney2024ai_cas}. Gudi{\~n}o et al.\ similarly argue that LLMs can function as agents in augmented democracy by processing public preferences at scales that are otherwise difficult to manage \cite{gudino}. Jarrett et al.\ formalize the idea of language agents as digital representatives that can stand in for individuals in collective decision-making \cite{jarrett2025languageagentsdigitalrepresentatives}. Together, this literature suggests that AI can play a more substantive role in democratic processes than recommendation or summarization alone.

Agora contributes to this emerging space, but differs from systems that use AI to directly synthesize collective outputs or simulate abstract publics. Instead, it uses AI to organize and operationalize a corpus of authentic human interviews: predicting support for novel policy proposals, surfacing the reasoning behind those stances, and allowing users to iteratively explore how policy changes shift support across a population. In this sense, Agora acts as a scaffold for learning how consensus-finding works under conditions of genuine disagreement.

\subsection{HCI and Deliberation}
The HCI community has long explored interfaces for public deliberation. Early systems like Opinion Space showed that making opinion spectra visible and explorable can increase respect for opposing views \cite{opinion_space}, a principle Agora builds on.

Systems like Procid scaffolded group deliberation by structuring engagement with divergent viewpoints to support consensus-building \cite{zilouchian2015procid}, underscoring the importance of balanced exposure to diverse perspectives in tools aimed at building deliberative capacity.
A parallel thread of HCI emphasizes structured reflection to improve deliberation quality \cite{yeo2025enhancing, yeo2024helpmereflect}. Prompts to articulate pros/cons and engage with others’ reasoning can foster deeper engagement \cite{considerit}, and reflection nudges, like clarifying views or adopting an opponent’s perspective, can increase attitude certainty and willingness to express opinions beyond what information access alone achieves \cite{reflection_nudge}.
Recent work also uses LLM-based critical thinking tools to improve users' argument quality and open-mindedness in online deliberation \cite{pan2025amquestioner, khadar2025wisdom}. These findings informed our design: we encourage users to engage with others’ lived experiences and reasoning to build broader consensus.
Prior work also shows that \textit{who} is speaking matters for learning. Platforms that surface opinions without stakeholder context can obscure lived consequences and intra-group variation \cite{PolicyScape}. Listening-centered designs can increase communication satisfaction and willingness to participate \cite{kriplean2012you}, and audio humanizes speakers in polarized contexts \cite{schroeder2017humanizing}. Accordingly, Agora pairs real profiles with voice clips to preserve tone and help users understand the reasoning behind different positions.


\section{Tool Description}

Agora is built on a foundation of voice-based interviews conducted with participants who shared their lived experiences and beliefs related to policy issues. We process these interviews with LLMs to (1) predict each interviewee's level of support for a given policy and (2) create audio medleys of experiences and reasoning that support those predictions.

\subsection{AI-Conducted Interviews}
\label{sec:ai-interview}

Agora is grounded in a corpus of semi-structured voice interviews conducted with 90 participants on Prolific. Participants were required to have at least one year of work experience and be currently living in the United States. We sought political balance by using Prolific's sampling settings to recruit a mix of liberal, moderate, and conservative participants.

We adapted the AI interviewer system from Park et al.  \citep{park2024generativeagentsimulations1000}, prioritizing low latency and voice-to-voice interaction to give participants the feeling of actually conversing with an interviewer (Appendix C). The system uses OpenAI's Whisper model for speech-to-text transcription~\cite{radford2023robust}, GPT-4o for generating contextually appropriate responses, and OpenAI's tts-1 model for text-to-speech synthesis. The system automatically detects pauses from the participant, transcribes their speech, and generates and vocalizes the next interviewer utterance. 
Like a semi-structured interview, the LLM begins with general background questions and follows up with curiosity and respect about the interviewee’s life. It then shifts to the two policy topics—minimum wage and domestic vs. foreign hiring—eliciting both personal experiences and beliefs (e.g., ``Have you or someone close to you ever been impacted by immigration policy, especially around hiring?'', ``How do you feel about companies prioritizing hiring local applicants over foreign applicants?''). Capturing both enabled us to represent diverse perspectives grounded in lived experience. The AI interviewer’s questions can be adjusted to align with the specific policy topic selected for display in the main user interface below.

\begin{figure*}
    \centering
    \includegraphics[width=1\linewidth]{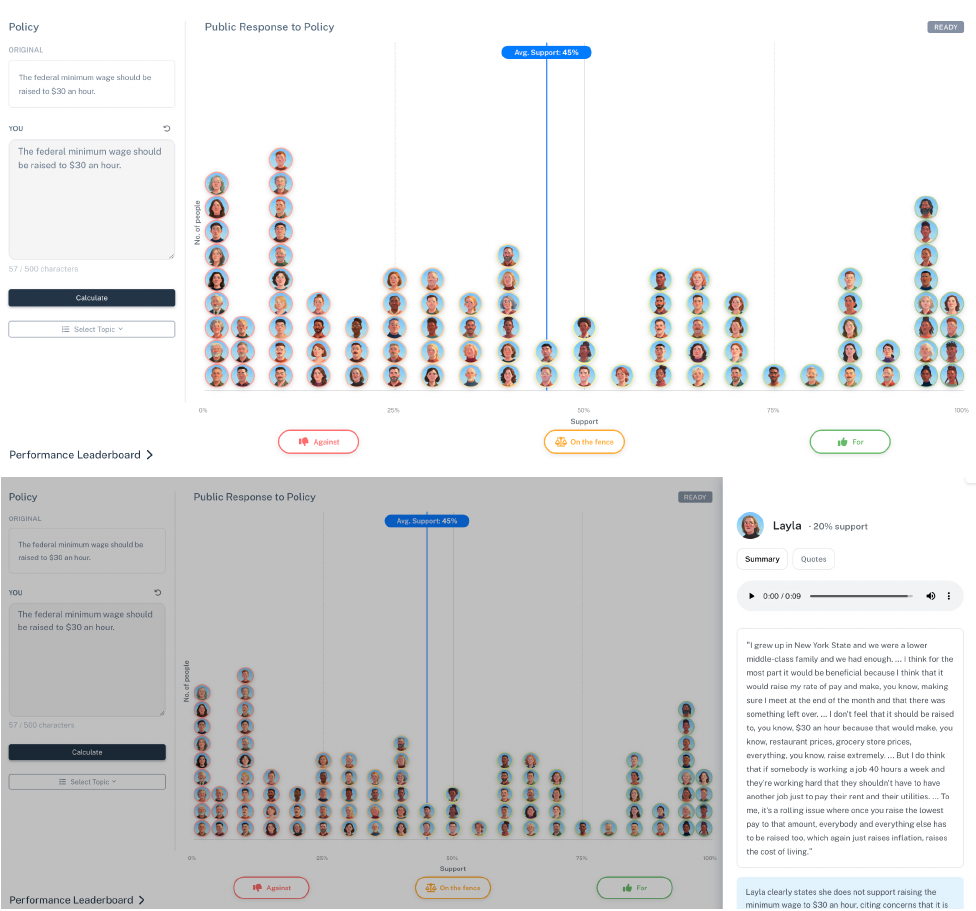}
    \caption[Main Agora interface (dynamic version)]{Full Agora interface for treatment condition. Top image A shows how participants iterate and test their policies, Bottom image B shows the profile view participants see when clicking on the different avatars. }
    \label{fig:main_agora_dynamic}
    \Description{Screenshot of the full Agora interface in the treatment condition. The top panel shows a policy input area on the left with text proposing to raise the federal minimum wage to \$30 per hour, alongside a “Calculate” button. The center displays a horizontal distribution of circular avatar icons depicting participant demographics positioned along a 0–100\% support scale, with a vertical line marking average support at 45\%. Avatars are grouped under “Against,” “On the fence,” and “For.” The bottom panel shows the same visualization with a side profile view opened for an avatar named Layla (20\% support), including tabs for summary and quotes, an audio playback bar showing the medley, and a text explanation of her stance.}
\end{figure*}

\subsection{Agora User Interface}

The interface (\autoref{fig:main_agora_dynamic}A), presents users with a policy-drafting task. On the left, users can write and revise their own policies. On the right, a visualization displays avatars representing the interviewees, positioned along a horizontal axis by their predicted support for the user's current policy (from 0\% to 100\%). The avatars were created with GPT-5 using the demographic information of each interviewee (age, race, and gender).
When users click on an individual avatar, they can listen to a 60--90 second audio medley of that person explaining their perspective in their own voice, along with a text summary (\autoref{fig:main_agora_dynamic}B). This medley presents experiences and reasoning supporting the model’s predicted stance, helping users understand not just where people stand but \textit{why} they hold their views. A leaderboard encourages users to iteratively maximize overall policy support. The LLM prompts are in Appendix D.

\subsection{Backend Implementation}

\subsubsection{\textbf{Predicting Policy Support}}

To position avatars along the support spectrum, we use GPT-4.1 to estimate each interviewee's level of support for a given policy. The model receives the interview transcript along with the policy text and is prompted to:
(1) Rate the predicted support of the interviewee from 0--100, 
(2) Provide reasoning for its prediction, and 
(3) Give a confidence score in its prediction.
While we did not use the model's reasoning directly in the interface, prompting for reasoning prior to giving an answer has been shown to increase prediction accuracy \cite{wei2023chainofthoughtpromptingelicitsreasoning}. 
We collected our interviewees' votes for the two policies—raising the federal minimum wage to $\$30$/hour, and whether companies should prioritize domestic over foreign applicants—via a pre-survey prior to interaction with the AI interviewer. 
We then validated the LLM predictions against these initial stances reported by interviewees, finding an average accuracy of 82\%.

\subsubsection{\textbf{Generating Audio Medleys}}

For each interviewee and policy proposal, we use GPT-4.1 to identify segments from the interview transcript that contain experiences and reasoning supporting the predicted stance. The model selects audio clips that ground the prediction in the interviewee's own words, prioritizing concrete personal experiences in line with research showing that such experiences are particularly effective at bridging moral and political divides~\cite{kubin, kessler2023hearing}. These clips are assembled into a 60--90 second medley that tool users can listen to when clicking on an avatar. 
A ``meta-medley'' feature uses GPT-4.1 to curate relevant clips from interviewees with low, medium, or high support, offering a quick cross-spectrum overview of perspectives without clicking through individual profiles. 

\subsubsection{\textbf{Dynamic Feedback Loop}}

A key feature is the dynamic feedback loop: when users revise their policy text and click ``Calculate,'' the system re-processes all interviewee transcripts through the LLM pipeline: 
(1) New policy text is sent to GPT-4.1 along with each interviewee's transcript.
(2) Support predictions are regenerated for all interviewees based on the updated policy.
(3) New audio medleys are generated that include experiences and reasoning relevant to the revised proposal.
(4) Avatar positions shift along the horizontal axis to reflect updated support.
(5) Individual profile views update with the newly generated medleys and summaries.
Users can immediately see how different policy framings affect the support distribution across the population. For instance, a user might discover that adding state-specific considerations to a minimum wage proposal shifts several previously opposed interviewees toward support, and can click on those avatars to hear the specific experiences and reasons that make such exemptions resonate.

\section{User Study}

\subsection{Experimental Setup}

We evaluated the tool with 44 university students (39\% Asian, 29\% White, 18\% Black or African American, 14\% Hispanic or Latino; 90\% undergrad and 10\% grad students with an average age of 19 years, 52\% identified as female, 46\% as male, and 2\% as gender neutral) in the United States in a fully online study approved as exempt under our institution's Institutional Review Board (IRB). We recruited these students through mailing lists, messaging platforms, and word of mouth.
Participants were asked to draft optimal policies on two topics: what the minimum wage should be and whether companies should prioritize hiring domestic over foreign applicants. Participants spent approximately 30-45 minutes completing the task and received \$10 for participation, with an additional \$50 bonus awarded to those whose policy proposals received the most support.

\begin{figure}
    \centering
    \includegraphics[angle=0,width=1\linewidth]{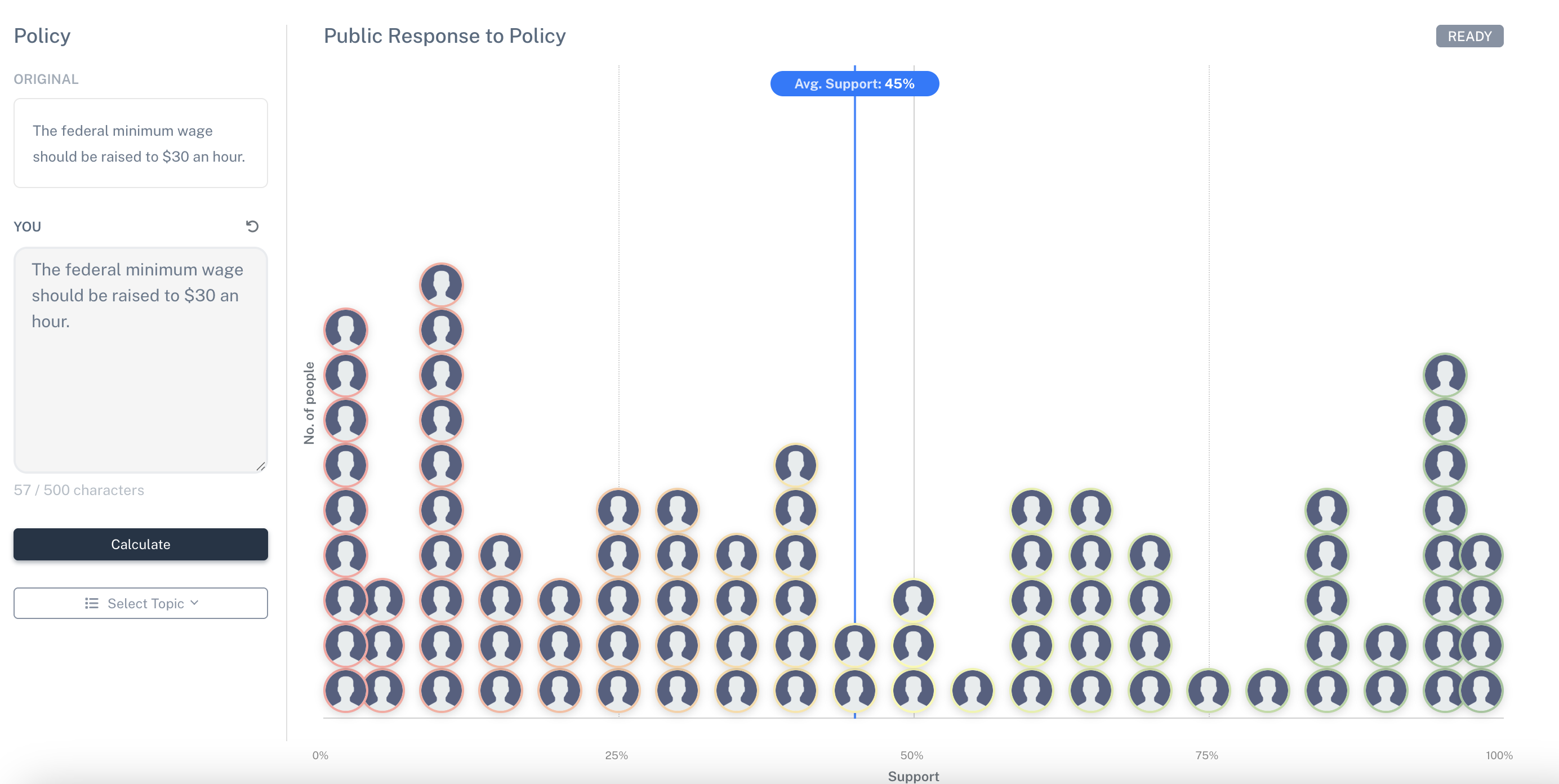}
    \caption{Agora interface for control condition}
    \label{fig:habermas-game-control}
    \Description{Screenshot of the full Agora interface in the control condition. The left panel shows a policy input area with text proposing to raise the federal minimum wage to \$30 per hour, along with a “Calculate” button and topic selection dropdown. The center displays a horizontal distribution of circular blank avatar icons positioned along a 0–100\% support scale, with a vertical line marking average support at 45\%. Avatars are clustered across the spectrum from low to high support. Unlike the treatment condition, no profile side panel, audio playback, summaries, or quote details are visible in this view.}
\end{figure}

During the task, participants aimed to maximize \textit{in silico} support from avatars grounded in the beliefs and experiences of the interview participants. Those in the treatment condition saw the full interface, enabling them to see and hear the reasons behind each avatar's predicted support, revise their policies, and observe how changes affected support levels. Those in the control condition used the same drafting tool, but avatars appeared as generic icons without interactive capabilities. Control participants could see \textit{how} the distribution of predicted support changed with new policies, but not \textit{why} it changed (see \autoref{fig:habermas-game-control}).
We chose this control design to isolate the impact of profile exploration and to establish a baseline for participants' prior knowledge and idea generation without external stimuli. Video walkthroughs for both conditions are in Appendix A.

\subsection{Data Collection}
\label{sec:data-collection}

\subsubsection{Self-reported Survey}
We evaluated the tool for its impact on participants' self-reported learning and the quality of the consensus statements via a 10-minute post survey (Appendix B3).
To evaluate learning impact, we adopted several subscales from \citet{Shroff_Ting_Lam_2019}, who designed and validated several scales measuring students’ perceptions of technology-enabled active learning. We used scales that measure (1) problem-solving skills, (2) interest, and (3) feedback. We also measured whether the tool helped participants (4) ``deliberate within.'' Deliberation within is a concept coined by \citet{Goodin2000}, for describing the process of considering and weighing alternatives by oneself prior to interpersonal deliberation. \citet{delib_within_measure} propose a scale to measure this concept, which we use to evaluate our tool.
Finally, we also added items assessing participants’ perceived (5) perspective-taking and empathy, including their ability to understand diverse viewpoints, interpret shifts in support, and recognize how others’ experiences and backgrounds shaped their reactions to the policy updates.

\subsubsection{Consensus Statement Data}
For subsequent quality analysis, all consensus statements submitted by participants were saved and labeled by participant ID, experimental condition, topic, and timestamp (to capture iterative progression of statements).

\subsubsection{Qualitative Interviews}
The first four authors conducted 12 semi-structured Zoom interviews (20–30 minutes each) across the two conditions, recruited via post-survey opt-in (with a \$10 incentive) and continued until saturation. We explored participants’ experiences encountering opposing views, learning from policy feedback, and attempting to build consensus. We specifically probed how their ideas evolved over moments of convergence, compromise, or reassessment. Finally, we examined perceptions of the interface, audio narration, and AI-generated summaries, focusing on authenticity, fairness, and impact on engagement and trust.

\subsection{Data Analysis}

\subsubsection{Survey Evaluation}
We compared survey responses between the treatment and control conditions across the five subscales described in \ref{sec:data-collection}. For each participant, we computed the mean score across items within each subscale. We then conducted Welch’s two-sample t-tests for each subscale and adjusted the resulting p-values using the Benjamini–Hochberg false discovery rate procedure.

\subsubsection{Consensus Statement Quality}

\paragraph{Rubric for Consensus Quality}

To evaluate consensus statement quality, we developed a rubric drawing on two established frameworks: VanSickle-Ward's statute specificity coding scheme \cite{VanSickleWardSpecificity2010} and Steenbergen et al.'s Discourse Quality Index (DQI) \cite{SteenbergenDQI2003}. These frameworks capture what makes a consensus statement useful rather than merely agreeable: it should be concrete enough to act on, and should reflect genuine engagement with competing perspectives rather than bare preference assertion \cite{Maia_2024, shaffer_delib}. 

The rubric, seen in Appendix E, assigns each statement a composite quality score based on four dimensions: validity, specificity, justification, and perspective acknowledgment. \textbf{Validity (0–1)} filters out irrelevant or adversarial submissions (e.g., prompt injection), assigning 0 to invalid entries and 1 to valid, good-faith policy statements.\textbf{ Specificity (0–3)}, adapted and abridged from VanSickle-Ward, captures whether a statement offers no clear stance (0), a general direction (1), a specified parameter (2), or multiple, interdependent parameters (3) \cite{VanSickleWardSpecificity2010}. \textbf{Justification (0–2)} and \textbf{perspective acknowledgment (0–1)} are adapted from the DQI, measuring inferential completeness and engagement with opposing positions respectively \cite{SteenbergenDQI2003}. The composite quality score is:
\begin{equation}
Q = V \cdot \left[\frac{1}{3} \cdot \frac{S}{3} + \frac{1}{3} \cdot \frac{J}{2} + \frac{1}{3} \cdot \frac{P}{1}\right]
\end{equation}
$V$ denotes validity, $S$ specificity, $J$ justification, and $P$ perspective acknowledgment. Validity is a multiplier that acts as a filter: invalid submissions ($V = 0$) receive a composite score of zero but are retained in the dataset. Specificity, justification, and perspective acknowledgment are normalized by their respective maximum scores and equally weighted, as we did not find a strong theoretical basis to prioritize one dimension over another.


To validate LLM-based scoring, we followed the LLM-as-a-judge paradigm, which uses language models to evaluate open-ended text against a rubric and has been shown to correlate well with human judgment across a range of tasks \cite{gu2025surveyllmasajudge}. Two human coders independently scored 12 pilot statements and revised the rubric instructions after discussing disagreements, and repeated the process twice to reach convergence. To assess intra-model consistency, we then ran Claude Opus 4.6 on the same 12 statements three times using the identical rubric; scores were highly consistent across runs. 
Cohen's kappa between the human coders' and Claude's median scores was 0.90 for specificity, 0.89 for justification, 0.83 for perspective acknowledgment, and 0.82 overall, indicating strong reliability for LLM-based scoring of the full dataset.

\paragraph{Quality Comparison} We compared consensus quality between the treatment and control conditions across the four dimensions and the composite quality score from the rubric above. For each participant, we first averaged the coded quality dimensions across all edited iterations of that recommendation topic. 
We compared the two conditions with two-sided Mann–Whitney U tests applied to the participant’s iteration-mean score for each dimension. Tests were run separately for each topic and again for pooled analysis, with Benjamini–Hochberg adjustment applied to control the false discovery rate across all tests. 

We also visualized the trajectories of consensus statement quality and support, comparing conditions to examine how these evolved as participants refined their statements over multiple attempts.

Finally, although not part of our original research focus, we supplemented our analysis by comparing the quality of consensus statements across the two recommendations to examine how topic familiarity may have influenced the results. For descriptive comparison of the two topics within each condition, we used two-sided Wilcoxon signed-rank tests on paired participant-level iteration-mean scores among participants with data for both recommendations, and adjusted Benjamini–Hochberg FDR.

\subsubsection{Qualitative Interviews Analysis}

We transcribed and cleaned all 12 interviews, then conducted an inductive thematic analysis, allowing patterns to emerge from the data \cite{charmaz2008grounded, corbin1990grounded}. The lead author first reviewed each transcript to develop notes, followed by a second pass to generate initial codes that were iteratively refined through discussions with the research team. We then grouped related excerpts into higher-level categories and themes, revising their definitions over four rounds until consensus and thematic saturation were reached. Finally, we compared how themes varied across conditions, consulting the broader team to validate interpretations and using these insights to contextualize quantitative findings. A complete list of themes can be found in Appendix F.

\begin{figure*}
    \centering
    \begin{subfigure}{0.42\linewidth}
        \centering
        \includegraphics[width=\linewidth]{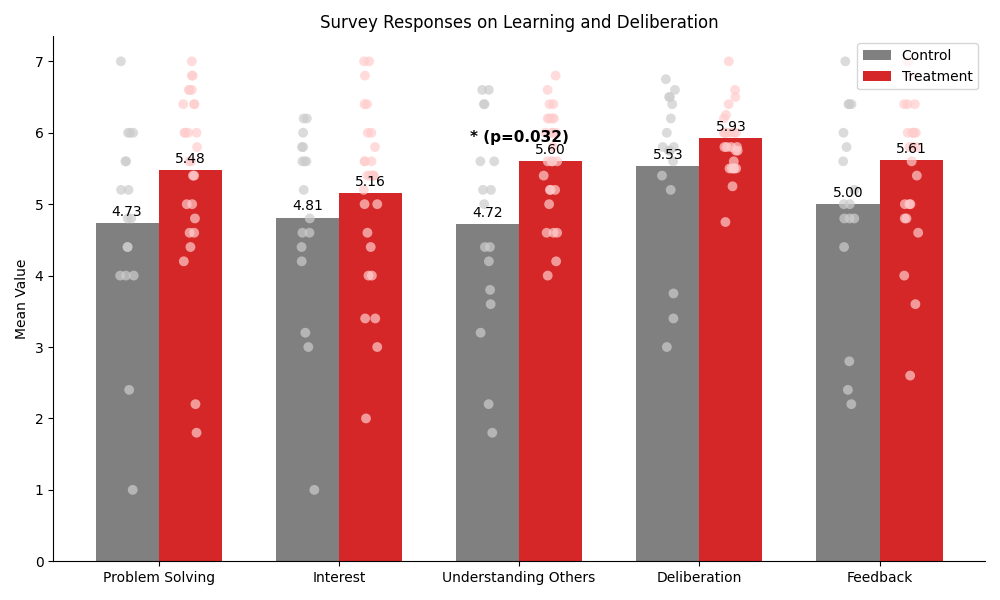}
        \caption{Mean survey responses across conditions measuring self-reports of learning and internal deliberation.}
        \label{fig:habermas-game-subjective}
        \Description{Bar chart titled "Survey Responses on Learning and Deliberation." The y-axis shows Mean Value from 0 to 7. Five measures are shown on the x-axis: Problem Solving, Interest, Understanding Others, Deliberation, and Feedback. For each measure, a gray bar represents the Control condition and a red bar represents the Treatment condition. Individual participant data points are overlaid on each bar as dots. Treatment scores are higher than Control scores across all five measures. The only statistically significant difference is on Understanding Others (Control: 4.72, Treatment: 5.60, marked with *p=0.032). Other mean values shown are: Problem Solving (Control: 4.73, Treatment: 5.48), Interest (Control: 4.81, Treatment: 5.16), Deliberation (Control: 5.53, Treatment: 5.93), and Feedback (Control: 5.00, Treatment: 5.61).}
    \end{subfigure}
    \hfill
    \begin{subfigure}{0.52\linewidth}
        \centering
        \includegraphics[width=\linewidth]{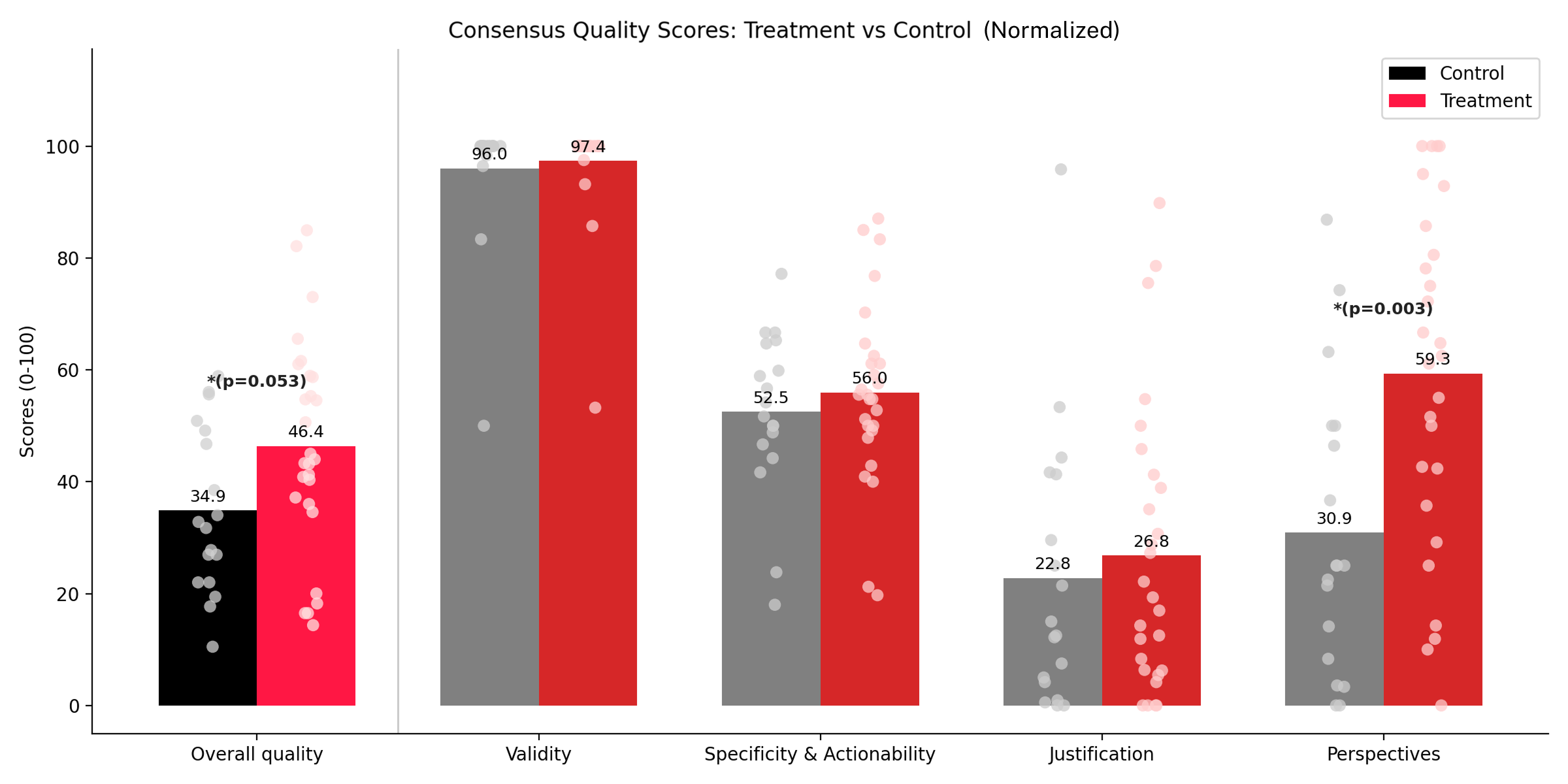}
        \caption{LLM-judged quality of consensus statements in treatment and control conditions (normalized).}
        \label{fig:habermas-game-quality}
    \end{subfigure}
    \caption{Agora learning outcomes and consensus quality.}
    \label{fig:habermas-game-combined}
    \Description{Bar chart titled "Consensus Quality Scores: Treatment vs Control (Normalized)." The y-axis shows Scores from 0 to 100. Five measures are shown on the x-axis: Overall Quality, Validity, Specificity & Actionability, Justification, and Perspectives. For each measure, a gray bar represents the Control condition and a red bar represents the Treatment condition. Individual participant data points are overlaid on each bar as dots. Treatment scores are higher than Control scores across all five measures. Two measures are marked with significance indicators: Overall Quality (Control: 34.9, Treatment: 46.4, p=0.053) and Perspectives (Control: 30.9, Treatment: 59.3, p=0.003). Other mean values shown are: Validity (Control: 96.0, Treatment: 97.4), Specificity & Actionability (Control: 52.5, Treatment: 56.0), and Justification (Control: 22.8, Treatment: 26.8).}
\end{figure*}

\begin{figure*}
    \centering
    \begin{subfigure}{0.48\linewidth}
        \centering
        \includegraphics[width=\linewidth]{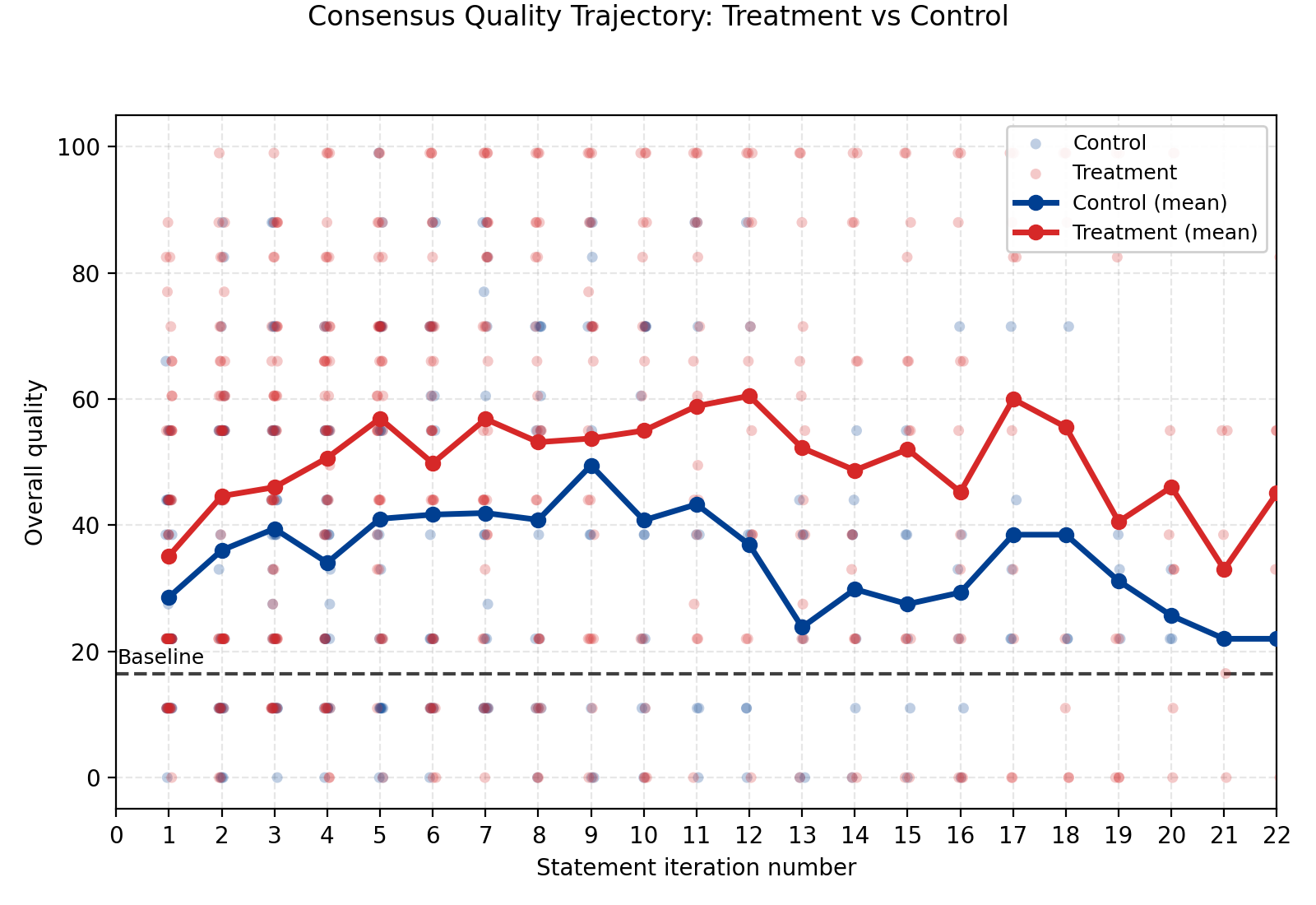}
        \caption{Comparison of consensus statement quality progression between the treatment and control conditions}
        \label{fig:trajectory_quality}
    \end{subfigure}
    \hfill
    \begin{subfigure}{0.49\linewidth}
        \centering
        \includegraphics[width=\linewidth]{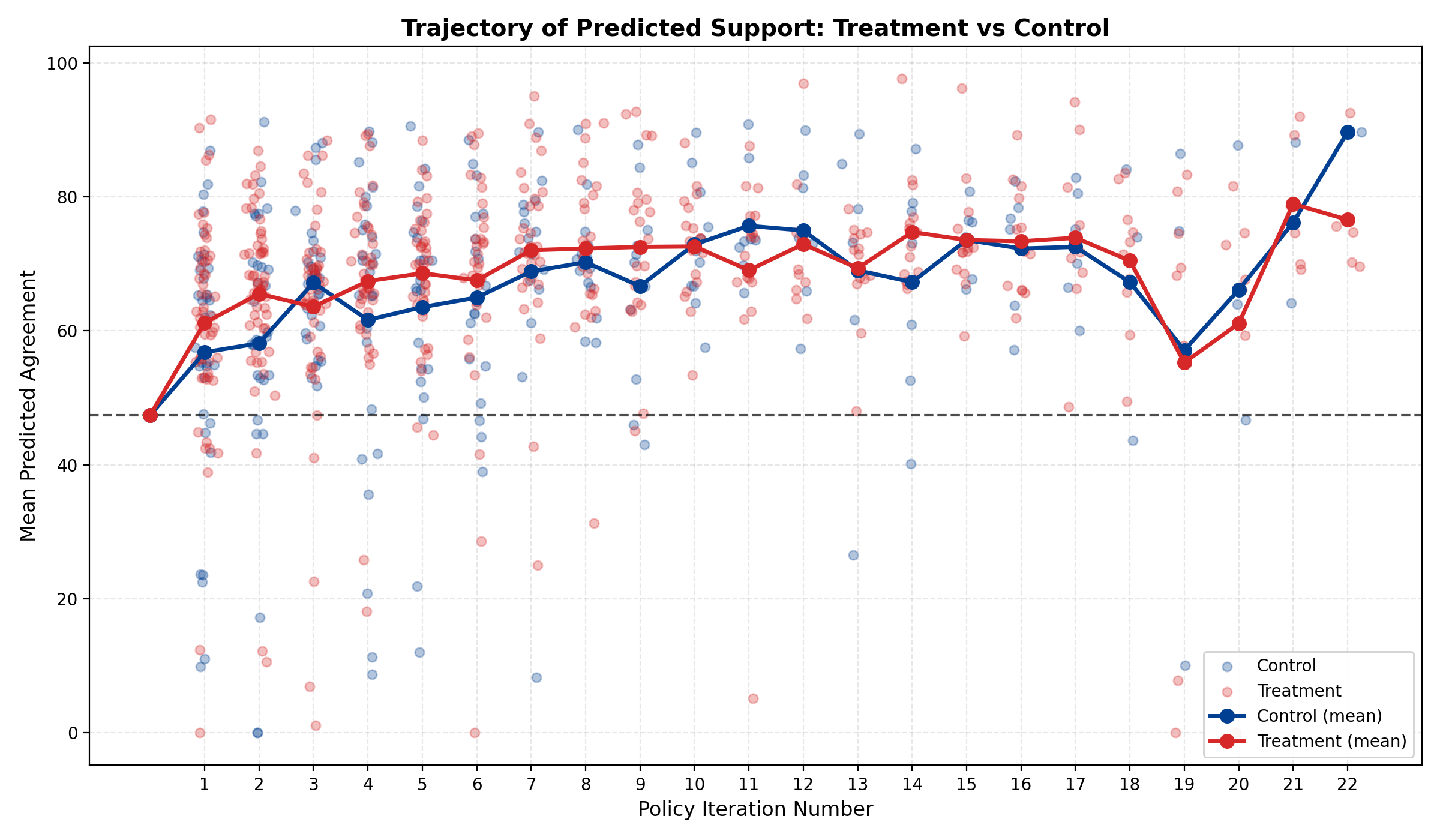}
        \caption{Comparison of predicted support trajectory between the treatment and control conditions}
        \label{fig:trajectory_support}
    \end{subfigure}
    \caption{Agora trajectory of statement support and quality}
    \label{fig:trajectory}
\end{figure*}

\begin{figure*}
    \centering
    \includegraphics[width=1\linewidth]{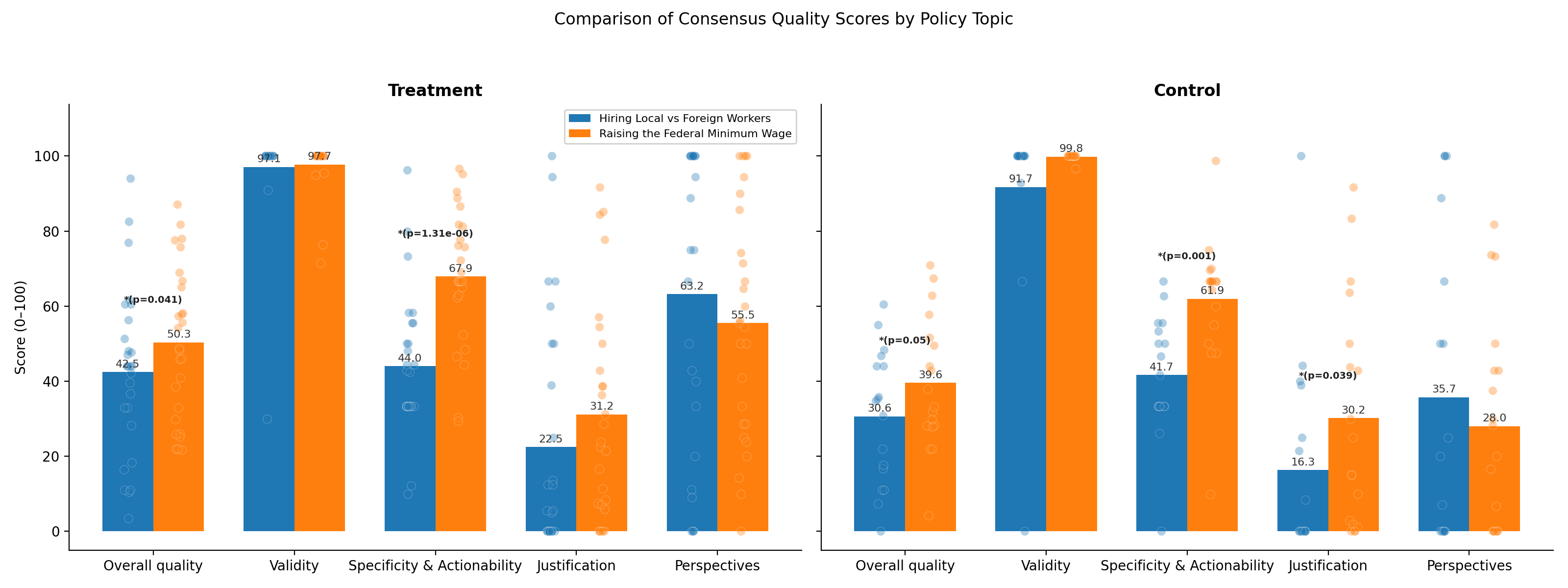}
    \caption{Impacts of policy topic on consensus quality scores for each condition }
    \label{fig:topic-comparison}
\end{figure*}

\section{ Results}

\subsection{Quantitative Results}


\subsubsection{Self-reports}
Participants in the treatment condition reported generally greater development of problem-solving skills, interest, and perceived the feedback from the tool as more timely and relevant (\autoref{fig:habermas-game-subjective}). They also reported higher scores on measures of ``deliberating within,'' suggesting that students felt the full tool was more effective at fostering internal deliberation compared to the control. However, we only obtained statistically significant differences for the ``understanding others'' subscale (5.60 vs 4.72, $d = 0.82$, $p = 0.032$). This likely reflects the treatment condition's exposure to diverse profiles, which may have encouraged greater empathy and social connection. Owing to the small sample size, these results warrant future investigation.

\subsubsection{Consensus Statements Quality}
Consensus statements produced by the treatment participants led to higher composite quality scores (\autoref{fig:habermas-game-quality}) than 
control participants (0.46 vs.\ 0.35, $r=0.35$, 
$p = 0.053$). Breaking this down, perspective acknowledgment 
was the only individual dimension to survive FDR correction (0.59 vs.\ 0.31, $r=0.52$, $p = .003$, $p_{\text{FDR}} = 0.01$). 
Validity (0.97 vs.\ 0.96, $r=0.07$, $p = 0.543$), specificity (1.68 vs.\ 1.57, $r=0.08$, 
$p = 0.635$), and justification (0.54 vs.\ 0.46, $r=0.09$, $p = 0.618$) trended 
higher in the treatment condition but did not reach statistical 
significance.

We also examined the progression of consensus statement quality as participants iterated on policies (\autoref{fig:trajectory_quality}). Treatment 
participants maintained consistently higher mean composite scores than control participants throughout the task (approximately 35--60 vs.\ 25--50 on a 0--100 scale). Here, we report trajectories up to iteration 22, beyond which fewer than 20\% of participants in either condition remained, making group averages susceptible to undue influence from a small number of individuals.

Both conditions achieved similar levels of predicted support by the end of the task, with mean predicted agreement converging to roughly 70--75\% for both groups after starting near the 50\% baseline (\autoref{fig:trajectory_support}). This pattern suggests that higher-quality statements do not necessarily attract greater predicted support: participants in the control condition were able to match the support levels of treatment participants through iteration, but produced lower-quality statements to do so. Quality and popularity, in other words, were not equivalent.

By topic (\autoref{fig:topic-comparison}), participants produced higher quality consensus statements on minimum wage than on domestic vs.\ foreign hiring priorities topic across nearly every dimension in both conditions. These differences were statistically significant for the overall quality score in treatment (50.3 vs. 42.5, $r=0.39$, $p = 0.041$) and control (39.6 vs. 30.6, $r=0.48$, $p = 0.05$). Participants also produced statements with higher statistically significant specificity and actionability for the minimum wage topic in both conditions (treatment: 67.9 vs. 44.0, $r=0.81$, $p = <0.001$; control: 61.9 vs. 41.7, $r=0.79$ , $p = 0.001$).

We triangulate these trends with findings from our qualitative interviews below.

\subsection{Qualitative Results}
Appendix F presents the complete set of categories, themes, and definitions derived from the interviews. In this paper, we focus on the starred (*) themes to contextualize the quantitative results, elaborating on them across the three subsections below.

\subsubsection{\textbf{Perspective Recognition and Engagement}}

\paragraph{\textbf{Perspective-Taking and Empathy}}
Treatment participants consistently reported that engaging with avatar profiles shifted their understanding of opposing viewpoints from positional disagreement to attributing others' stances to geographic origin, occupational history, educational background, and personal economic circumstance. Crucially, participants extended genuine engagement to views anchored in lived experience while dismissing those perceived as arbitrary: \textit{``Once I started listening to their feedback and their own stories, I was like, oh, this makes sense. What I want is actual reasoning behind people's beliefs. [The] moment they're like, `no [...] they don't work hard enough,' that's just unreasonable'' (T4).} T6 similarly noted that \textit{``having the backstory for each avatar was really helpful, it helps you consider why they're coming to the conclusion that they're coming to.''} Several participants explicitly connected these details to justified reactions to their policy revisions; T6 noted that voice recordings helped \textit{``contextualize the changes that would be made if this policy went into effect,''}.
The audio medleys also produced moments of unexpected affective connection that went beyond strategic perspective-taking. T3 noted that \textit{``seeing text on a page is not hearing a person...you are talking to a person and not an opinion''}, likening the experience to \textit{``sitting on a porch next to the person in rocking chairs.''} T4 described recognizing a shared identity with an avatar who held an opposing view: \textit{``there was this one person who said, `I come from a rural area, middle of nowhere, and I'm struggling.' I'm like, that's literally me—it started off with something we can share instead of just the AI summary that says, `I oppose.'''} 

Without access to individual profiles, control participants however were left to reason about distributions rather than people, and their accounts revealed a persistent awareness of what was absent. C1 reflected: \textit{``I wasn't making any progress, and that's when I found that `why are you voting the way you are?' was kind of coming to mind more, because I didn't know enough of the issue to be able to estimate.''} C3 noted \textit{``roughly how split we are, but not why one side is thinking the other.''} As a result, they reported bypassing perspective-taking altogether, instead trying to make their statement \textit{``as neutral as possible by adding more and more words that kind of mask the meaning of the statement itself''} to garner broader support. Control participants thus treated the process as a purely technical exercise: \textit{``it feels like I'm a data analyst trying to maximize my numbers, but if you have those numbers become people again, I can see why you are disagreeing, and I can think more carefully about how to compromise between two people, rather than just maximize the number.'' }

\paragraph{\textbf{Consideration of Own Viewpoints}}
A consistent pattern among treatment participants was the deliberate effort to step back from their own positions and adopt something closer to a third-party stance: an orientation made possible, and in many cases actively prompted, by the availability of individual profiles to reason from. T3, who entered the task \textit{``very pro-minimum wage,''} redirected their attention: \textit{``I didn't enter it wanting my opinions to be heard, I was trying to enter it as a third-party sort of policymaker. I need to have an open mind to everybody's concerns.''} In practice, this meant reading profiles systematically and doing \textit{``tally marks on what are the biggest concerns and what should I target in the policy.''} T4 reflected that encountering the full distribution recalibrated their priors: \textit{``I didn't realize \$20 was that controversial... maybe my opinion has a lot of haters—we're all stuck in our own bubble.''} 

Without profile access to reason from, control participants had no external anchor and fell back on personal knowledge, inference, and ideological imagination: \textit{``I used just meta-knowledge, I'm vaguely aware that people want to make sure the market doesn't feel bad effects.'' (C1).} This often ran up against limits of domain knowledge: \textit{``I didn't know enough of the issue [domestic vs foreign hiring] to be able to estimate,''} while C3 similarly noted that ``trying to design a policy without that background is a little bit difficult.'' The most striking contrast came on the question of policy revision: C2 reported, \textit{``I don't think I did really consider other points of view...for me personally, these were the best policies I could think of, and I still stood by it.''} 

\subsubsection{\textbf{Learning about Policies and Consensus-building}}

\paragraph{\textbf{Policy Complexity:}}

Engagement with avatars surfaced policy tradeoffs that treatment participants reported not having previously considered: \textit{``I always thought minimum wage was set at the federal level—I didn't know certain states had higher minimum wages, so seeing region-specific people was very interesting'' (T7).} Most participants also noted being less familiar with the domestic versus foreign hiring topic than minimum wage. As T4 noted: \textit{``I was learning more during the globalization one—I know a lot about minimum wage... But globalization, I was like, you're just offshoring your workers, there's no point. And then a lot of people are saying, `no, it's because other countries have more specializations, there are reasons to hire people who are not in the US.'''}  T3 reported a revelation on the hiring topic: \textit{``some people were like, `racism is bad,' and others were thinking we should prioritize our communities—and I'm thinking, wait, these are not mutually exclusive. I had never heard that argument before. Not a problem I had considered.''}

Control participants also encountered these challenges but were unable to resolve them: C1 reported being `\textit{`forced to reckon with the fact that there is difficulty''} around inflation and market effects; on the hiring topic they acknowledged that they \textit{``just weren't aware of the nuances of both sides''} to learn or solidify their viewpoints on the topic further.

\paragraph{\textbf{Rhetoric and Framing as Civic Skill:}}

Treatment and control participants converged on a shared insight—that how a policy is worded matters as much as what it proposes—but arrived there through qualitatively different routes. Treatment participants adjusted language in direct response to specific concerns surfaced by individual profiles, making framing a byproduct of perspective engagement. T1 changed emphasis \textit{``depending on what they said they valued or what was really important to them,''} and T3 systematically compared policy iterations: \textit{``I'd change the wording of something to make it seem stronger, or add more things if I'm seeing in the feedback that maybe they want more support for small businesses.'' } T2 described surprise at how much surface language moved support: \textit{``the way you say something can really change people's willingness to operate and engage with [it]... I'll try to tinker with a word here or a word there.''} T9 drew an explicit civic lesson from this pattern, observing that \textit{``the less you say, it seems better.''}

In the control condition, framing manipulation was the primary, and often, only tool available, making it less a civic insight than a mechanical strategy: \textit{``I didn't try to consider as many points of view so much as trying to convolute the statement''} (C3). C1 similarly discovered the effects of length and tone through trial and error without being able to connect those to concerns of specific individuals. Where treatment participants learned framing matters because different people weigh language differently for reasons rooted in their lives, control participants learned that framing matters as a numeric fact, decoupled from the human reasoning behind it.

\paragraph{\textbf{Consensus Conceptualization}}
Treatment participants rarely held a stable, singular definition of consensus, and their accounts reveal a productive tension between score maximization and genuine compromise that evolved over the course of the task. T5 initially defined finding consensus simply as \textit{``getting as high a score as possible on the approval rating,''} but the act of reading profiles pushed that framing toward something more substantive. T4 described a mid-task shift: \textit{``at the beginning I was just like, I gotta get the highest, and then at some point my brain shifted to—I need to see if I can make an understanding between these two groups. It clicked that half of us want one thing and half want the same thing, just not as much.''} T2 framed consensus explicitly in civic rather than numerical terms: \textit{``what can I do to get as many people to support a measure that I think is gonna help their lives—finding a way to word it and build it in a way that gets as many hands together as possible.''} T6 responded to avatar concerns about skilled labor parity by constructing a tiered wage policy: raising the general minimum while setting a higher floor for first responders, while T4, on the hiring topic, proposed that overseas workers be hired only when demonstrably more skilled, and otherwise training domestic workers. T5 concluded consensus was achievable \textit{``so long as you defined it as the majority having more than 75\% satisfaction, if you tried to define it as everyone being fully happy, there were always going to be outliers.''}

Control participants converged on a leaner and more static conception. C3 was explicit: \textit{``consensus means straight numbers—trying to get as much support as possible and minimize the divide in the middle.''} Without profiles to refer to, compromise was largely numerical—sliding a figure down, softening language—rather than substantive. C1's iteration consisted of \textit{``playing around with the most extreme versions of the bill and then finding the best compromise''} rather than a response to identifiable people. 

\subsubsection{\textbf{Model Transparency and Real-world Applications}}

In both conditions, participants expressed a persistent desire to understand why the scoring model responded the way it did, touching on the credibility of the tool itself. The most common frustration was the gap between a targeted policy addition and an unexpected or flat change in support: T4 described adding an explicit subsidy for small businesses and finding the model ``acted like it never existed.'' For some, the opacity extended to a deeper skepticism about the model's fidelity. T4 questioned whether a percentage score adequately captured human nuance: \textit{``I think people are more nuanced than saying I'm 80\% sure of this''}, while T9 suggested that publicly explaining the modeling process would itself be a form of legitimacy: \textit{``perhaps explaining we interviewed people on pertinent concerns, and each one represents a certain amount of voters: that would make it more credible.''} T2 framed transparency as a civic requirement: `\textit{`I would love to get a little bit of understanding of the black box behind that and I think that would be very important if you were going to scale this tool up, because people are going to be more and more distrustful of AI.''} 

Several participants imagined the tool operating beyond a research or educational setting. T2 saw potential in feeding Agora's outputs into actual legislative offices: \textit{``representatives have aides, they have people that work with them, so if there was a way to get this tool in a space that could feed this information to representatives, that would feel a lot more real''} and describing how the stakes of engagement would change entirely if the tool could influence real material outcomes. T9, drawing on a background in economics, proposed a set of extensions oriented toward electoral realism: modeling likely voter turnout alongside approval, grouping avatars by demographic and ideological clusters, and incorporating a primary-to-general-election arc to simulate different voter pools that real candidates face. T6 noted that instantaneous feedback, while pedagogically useful, \textit{``was a reminder that they are not in a real-life setting...people voting would be a much longer process''}, suggesting that future versions might use delayed or batched feedback to better simulate actual democratic rhythms.

\section{Discussion}

Our findings offer initial evidence that AI can scaffold consensus-finding as a learnable civic skill—and that the key may be access to the experiences behind drive others' views, not just the views themselves.

\textbf{Self Reports.} 
The only self-report direct measure to achieve statistical significance—understanding of others' perspectives—is the dimension most directly targeted by Agora's design. Participants in the treatment condition, who could explore the reasoning and lived experiences behind each avatar's predicted stance, rated their perspective-taking significantly higher than those who saw only aggregate support distributions. Taken together, this is consistent with deliberative pedagogy's emphasis on the depth, not just the presence, of engagement with difference \cite{Maia_2024, McDevitt01072006, shaffer_delib}. It also aligns with prior evidence that giving disagreement a concrete, navigable form encourages engagement with opposing views \cite{considerit}.

\textbf{Perspective Acknowledgment.} While overall consensus quality trended higher, with borderline significance, in the treatment condition, the strongest individual effect was on perspective acknowledgment. This was the dimension most directly tied to what treatment participants had access to that control participants did not: individual stories and voices. Here, there are three plausible and complementary mechanisms. 

The first is structural: treatment participants were incentivized to maximize predicted support, which required attending to the specific concerns of individual avatars. As (T6) described, they would ``interact with each of the little avatars, both for and not in favor, and then try to adapt [their policy] to make sure some of the people who are not in favor... their experiences were reflected.'' This is the deliberative skill Agora is designed to facilitate—iteratively revising a proposal while tracking whose concerns it addresses. Prior systems have produced similar effects: Kriplean et al.'s Reflect system \cite{kriplean2012you} found that prompting users to actively restate other's points—not just read them—increased communication satisfaction, in part by forcing genuine listening. Agora embodies a similar principle in its incentive design, compelling participants to incorporate feedback from individual avatars to maximize support.

The second mechanism is affective: hearing real voices increased participants' motivation to engage with opposing views. One participant reflected, ``it's not just more money, it's because they have experience with having this amount, and they know it's not enough'' (T4) — suggesting a shift from treating avatars as preference aggregates to treating them as people with reasons. This aligns with Schroeder et al.'s finding that voice humanizes speakers in polarized contexts \cite{schroeder2017humanizing} and with Kubin et al.'s finding that personal experiences bridge moral divides more effectively than abstract arguments \cite{kubin}. 
That treatment participants rated their "understanding of others' perspectives" significantly higher supports both accounts: the tool encouraged perspective-taking both by design and through the humanizing effect of authentic voice. 

The interviews revealed a third mechanism. Viewing profiles prompted treatment participants to consciously inhabit a third-party stance—acting as a "policymaker with an open mind to everybody's concerns" (T3)—distinct from both the structural incentive and the affective response to voice. Control participants, lacking profiles, remained anchored in prior beliefs or resorted to hypothetical inferences, consistent with Kim et al.'s finding that opinions without stakeholder context leave users reasoning about aggregate positions rather than the lived experiences behind them \cite{PolicyScape}.
Together, our findings suggest that perspective acknowledgment in the treatment condition was driven not only by what participants heard, but by the role the interface implicitly assigned them.


\textbf{Specificity and justification} trended in the expected direction but did not reach significance. We believe this reflects a meaningful tension in the tool's design. Participants were instructed to maximize predicted support from avatars, but this score did not inherently reward more specific or better-justified policies. In practice, the incentive structure could push participants in the opposite direction: participants in the interviews noted that vague or emotionally resonant framings sometimes garnered higher agreement especially in the control condition. A popular policy is not always a higher-quality one, and the current design conflates these. 

Our interviews showed that the treatment participants conceptualized consensus as finding overlap between genuinely held positions, while control participants maintained a predominantly numerical conception; this may help explain not only the perspective acknowledgment gap but also the directional (if non-significant) differences in justification. Treatment participants had the materials to justify each compromise; control participants did not. Some participants also observed that the one-dimensional support scores flattened the nuance of how individuals react to policy changes. Future versions of Agora can incorporate statement quality feedback alongside the predicted support score to better align the tool's incentive structure and feedback loop with its educational aims.


\textbf{Topic Variability.} Participants produced higher quality consensus statements on minimum wage than on domestic vs. foreign hiring across nearly all dimensions and both conditions. This may partly reflect differences in topic familiarity: minimum wage may feature more prominently in everyday US political discourse, and our participants may have entered with stronger prior frameworks and experiences (e.g. regional variation in cost of living, downstream inflation effects) for reasoning about it. For the hiring topic, by contrast, profiles often introduced new conceptual frames that participants lacked the prior knowledge to integrate into coherent policy. The result was that, on the hiring topic, participants were learning the landscape of the problem at the same time as they were trying to craft consensus. This added cognitive load that may have suppressed the specificity and justification gains otherwise driven by profile engagement.
Future work should counterbalance topic order and integrate topic familiarity to disentangle these effects.

\textbf{Conclusion.} This paper makes three contributions to HCI. First, we introduce Agora, a system that treats consensus-finding as a learnable civic skill through iterative policy drafting, exposure to authentic perspectives, and immediate feedback on predicted support.
Second, we demonstrate a concrete use case for LLMs in democratic contexts that does not replace human voice: by using LLMs to organize a corpus of real interviews rather than simulating personas, Agora grounds its feedback in authentic human stories. This contrasts with systems such as Tessler et al.'s Habermas Machine \cite{habermas}, where AI generates consensus statements directly. Agora users instead write their own policies while AI surfaces the perspectives that inform them—demonstrating a productive alternative design direction for democratic applications of LLMs. Third, we provide preliminary experimental evidence that access to voice-grounded explanations—as opposed to aggregate support distributions alone—significantly improves both self-reported perspective-taking and the quality of written consensus statements along the dimension of perspective acknowledgment. 

More broadly, our experimental findings suggest that the \textit{why} behind others' positions matters: understanding the reasons people hold their views, not just the distribution of those views, appears central to developing the perspective-taking capacities that deliberative theorists identify as foundational to democratic competence \cite{kirlin2003role}. If deliberative skills can be practiced through mediated engagement with authentic perspectives, tools like Agora could help address deliberative democracy's scalability challenge by widening access beyond the small number of citizens who participate in well-designed deliberative 
processes \cite{fishkin2009people}—serving not as a replacement for face-to-face deliberation, but as preparation for it.


\section{Limitations and Future Work:}
\textbf{Sample size and generalizability.} Our sample consisted of 44 university students recruited through convenience sampling, and the study was conducted fully online. University students likely enter with above average baseline deliberative exposure and political knowledge, which may have attenuated treatment effects or produced ceiling effects on specificity and justification. Results may not generalize to populations with lower prior engagement with these topics, less experience with digital interfaces, or stronger prior partisan commitments. Additionally, with 44 participants, the study was likely underpowered to detect smaller effects; findings warrant replication with a larger sample before stronger causal claims can be made. Future work should evaluate Agora with larger and more diverse samples, including participants with direct policy stakes. Future designs might also consider adaptive scaffolding that adjusts profile depth or guided prompts based on participants' domain familiarity. 

\textbf{Isolating variables. }The current control condition isolates the effect of profile exploration broadly, but cannot distinguish between the contributions of specific features: voice clips vs. text summaries, dynamic support feedback, or the iterative revision structure itself. Future work should vary access to these components independently to identify which elements drive deliberative learning outcomes.

\textbf{Incentive-quality misalignment.} As discussed above, the current design rewards predicted support, which is not equivalent to policy quality, and may actively work against the development of specificity and justification. Future iterations should explore whether quality-sensitive scoring, structured reflection prompts, or explicit feedback on argument completeness can better align the tool's feedback loop with its educational aims \cite{delib_within_measure, yeo2024helpmereflect, yeo2025enhancing}.

\textbf{Topic order and familiarity.} All participants completed the minimum wage task before the domestic versus foreign hiring task, and order was not counterbalanced. Differences in consensus quality across topics may reflect topic familiarity, participant fatigue, or both—these cannot be disentangled in the current design. Future work should counterbalance topic order and vary topic familiarity to separate these effects.

\textbf{Representing nuances in support.} The current visualization reduces policy support to a single axis, which enables comparison and mirrors how public opinion is often summarized, but obscures the multidimensional and heterogeneous nature of real-world preferences. A policy that appears broadly supported may rest on incompatible rationales. Future versions could explore richer representations like clustering avatars by reasoning philosophy rather than aggregate support. Participants also noted frustration at not being able to track whether the same individuals drove persistent low-support ratings. Here, enabling persistent profile tracking across iterations would make the interface more deliberative.

\textbf{Model transparency and credibility.} 
Given participants' skepticism of the function and credibility of Agora's predictive support scores, future iterations could explain the modeling process as part of onboarding: how interview data builds predictions, what confidence scores mean, and how policy text is matched to transcripts.

\textbf{Ethical concerns.} The tool does not fully address privacy (e.g., voice anonymization), and future versions should give interviewees  in \autoref{sec:ai-interview} more agency over disclosure. LLM biases may have shaped how perspectives were organized and support updated; while prompts are public, model behavior remains hard to interpret. Future work should solicit interviewees' feedback on how their views are represented and how support shifts are portrayed.

\textbf{Depth and transfer of learning.} We measured self-reported deliberative outcomes and single-session consensus statements. Whether gains persist or accumulate across sessions is unknown; longitudinal designs and behavioral measures beyond self-report can strengthen causal claims.


\textbf{Relationship to face-to-face deliberation.} Agora is designed as a training ground, not a replacement, for interpersonal deliberation. Like any mediated format, it may produce engagement that is real but shallower than face-to-face encounters. Future work should examine whether Agora improves subsequent face-to-face deliberation, and explore institutional applications through co-design with policymakers and civic educators.

\section{Acknowledgments}
We would like to thank the Foresight Institute for supporting this work.

\bibliographystyle{ACM-Reference-Format}

\bibliography{sample-base}

\appendix

\section{Platform Demo and Instructions}
The following videos were shared with participants prior to the study and also demonstrate how the tool works:
\begin{itemize}
    \item {Experimental Condition:} \href{https://drive.google.com/file/d/1z4irIiTkOmYEauwgg9Aiff8d3Qu_RY-5/view?usp=drive_link}{Video link}
    \item {Control Condition:} \href{https://drive.google.com/file/d/18uxrT4WK0sdOFo23aQqQmcH2nuKwa2cn/view?usp=drive_link}{Video link}
\end{itemize}

\section{Questionnaires}
\subsection{Pre-Activity}
\subsubsection{Demographics}
\begin{enumerate}
    \item What is your age?
    \item What is your gender identity?
    \item How would you describe your race?
\end{enumerate}

\subsection{Pre/Post-Activity}
\subsubsection{Policy Benefits and Drawbacks}
Participants were allowed to list up to 10 answers to each question.
\begin{enumerate}
    \item What are the potential benefits of raising the federal minimum wage? Write as many benefits as you can think of. Please enter at least one benefit to proceed. If you cannot think of any, you may write ‘idk.’
    \item What are the potential drawbacks of raising the federal minimum wage? Write as many drawbacks as you can think of. Please enter at least one drawback to proceed. If you cannot think of any, you may write ‘idk.’
    \item What are the potential benefits of companies prioritizing domestic versus foreign workers when making hiring decisions? Write as many benefits as you can think of. Please enter at least one benefit to proceed. If you cannot think of any, you may write ‘idk.’
    \item What are the potential drawbacks of companies prioritizing domestic versus foreign workers when making hiring decisions? Write as many drawbacks as you can think of. Please enter at least one drawback to proceed. If you cannot think of any, you may write ‘idk.’
\end{enumerate}

\subsection{Post-Activity}
\subsubsection{\textbf{Sub-scales from Technology-Enabled Active Learning Inventory (TEAL)}}

\paragraph{Adapted from Shroff et al.}
Participants rated their agreement with the following statements using a 7-point Likert scale (Strongly Disagree; Disagree; Slightly Disagree; Neither Agree nor Disagree; Slightly Agree; Moderately Agree; Strongly Agree).

\paragraph{Problem-Solving Skills (PRS)}
\begin{enumerate}
  \item This tool allowed me to methodically generate ideas by contributing information from multiple viewpoints.
  \item This tool enabled me to solve a problem systematically by taking into account different points of view.
  \item This tool encouraged me to think critically about the broader concepts related to the problem.
  \item This tool let me analyze my own views and their wider contexts in order to draw firm conclusions.
  \item This tool allowed me to define the problem systematically by viewing it from different angles in an effort to find possible solutions.
\end{enumerate}

\paragraph{Interest (INT)}
\begin{enumerate}
  \item This tool allowed me to engage in thought-provoking dialogue with points of view that challenged my perspectives.
  \item This tool encouraged me to explore a variety of different issues that I may not have otherwise considered.
  \item This tool piqued my curiosity by exploring various options when navigating the user interface.
  \item This tool held my attention by challenging me to look into issues that I may not have otherwise thought of.
  \item This tool encouraged me to exert effort in the face of difficulty by persisting at tasks I found challenging.
\end{enumerate}

\paragraph{Feedback (FEE)}
\begin{enumerate}
  \item This tool allowed me to receive timely feedback that helped me improve my performance.
  \item This tool enabled me to receive inputs that helped me keep track of my own performance.
  \item This tool allowed me to receive prompt feedback so that I was aware of my progression toward knowledge acquisition.
  \item This tool allowed me to receive prompt feedback so that I was aware of my progression toward mastery of my skills.
  \item This tool enabled me to receive responses that supported my further understanding.
\end{enumerate}

\subsubsection{\textbf{"Deliberation Within" Scale}}

\paragraph{Adapted from Weinmann et al.} Participants rated their agreement with the following statements using a 5-point Likert scale (Strongly Disagree; Disagree; Neither Agree nor Disagree; Agree; Strongly Agree).

\begin{enumerate}
    \item I have reassessed my biases favoring or opposing different solutions.
    \item After listening to the advice of others, I have taken responsibility for making up my own mind about the topic.
    \item I have reflected on several opinions about the topic.
    \item I have thoughts about arguments for and against my own as well as others' opinions about the topic.
    \item I have evaluated the arguments that speak for and against my own as well as for and against others' opinions.
\end{enumerate}

\subsubsection{\textbf{Custom Items}}
\paragraph{\textbf{Understanding}: }Participants rated their agreement with the following statements using a 7-point Likert scale (Strongly Disagree; Disagree; Slightly Disagree; Neither Agree nor Disagree; Slightly Agree; Moderately Agree; Strongly Agree).

\begin{enumerate}
    \item I understood why people with different opinions on this topic might feel the way they do.
    \item I had a strong understanding of how the opinions and experiences of the people in the sample population informed their responses to the proposals.
    \item I now have a better sense of how people think about this issue.
    \item I understand why my proposals received more or less support.
    \item I found it easy to understand how changes to my proposal affected its level of support.
\end{enumerate}

\paragraph{\textbf{User Experience}:}
Participants rated their agreement with the following statements using a 7-point Likert scale (Strongly Disagree; Disagree; Slightly Disagree; Neither Agree nor Disagree; Slightly Agree; Moderately Agree; Strongly Agree).

\begin{enumerate}
    \item The tool was easy to use.
    \item The tool was enjoyable to use.
    \item The tool was frustrating to use.
    \item I would like to use this tool again.
\end{enumerate}

\paragraph{\textbf{Qualitative}:}
Participants responded in a short text field.

\begin{enumerate}
    \item What did you like most about the experience? For example, what features did you like?
    \item What about the tool felt confusing or unnatural?
    \item How did this experience change how you think about finding consensus, if at all?
\end{enumerate}

\paragraph{\textbf{Live Interview Questions}:}
The following custom items were asked during optional participant interviews.

\paragraph{Perspective-taking}
\begin{enumerate}
  \item When you encountered viewpoints different from your own during the game, what stood out to you most?
  \item As you were writing the statements, how did you come up with or consider other points of view?
  \item Can you describe a moment when you began to understand why someone might hold a very different opinion?
  \item Did the game help you see how people’s backgrounds or experiences shape their reasoning on policy questions?
  \item How did the design of the tool (e.g., interface, audio, visual summaries) help or hinder your ability to take others’ perspectives seriously?
  \item Do you think your understanding of opposing views changed between the beginning and end of the experience? How so?
\end{enumerate}

\paragraph{Learning (knowledge gain and self-reflection)}
\begin{enumerate}
  \item What new information, concepts, or arguments did you learn through the game?
  \item Did you play any voice recordings? How did the prompts, medleys, or summaries influence how you thought about the policy issues?
  \item Did you feel that you were learning with others or competing against others?
  \item Looking back, did you reassess any of your prior assumptions? What prompted that reflection?
  \item Did you feel more confident in explaining both sides of the issue after playing the game?
\end{enumerate}

\paragraph{Consensus-building skills}
\begin{enumerate}
  \item What did ``finding consensus'' mean to you in this context?
  \item Did the game make consensus feel achievable or artificial? Why?
  \item How did you decide when a proposal or statement was ``good enough'' to be popular or well-reasoned?
  \item Did you notice how your ideas evolved over time? Or were there clear moments of convergence or compromise?
\end{enumerate}

\paragraph{Role of technology}
\begin{enumerate}
  \item How did the game’s interface and audio narration affect your engagement?
  \item Did the AI-generated medleys or summaries feel authentic and fair?
  \item Was it clear what the rules or goals of the game were?
  \item Did you experience any confusion or frustration with how your inputs were reflected or scored?
  \item What, to you, would make the game feel more like a meaningful deliberation rather than a simulation?
  \item In what ways did the design make you feel included or excluded from the process?
\end{enumerate}

\section{AI Interviewer Details}
\begin{figure}
    \centering
    \includegraphics[width=1\linewidth]{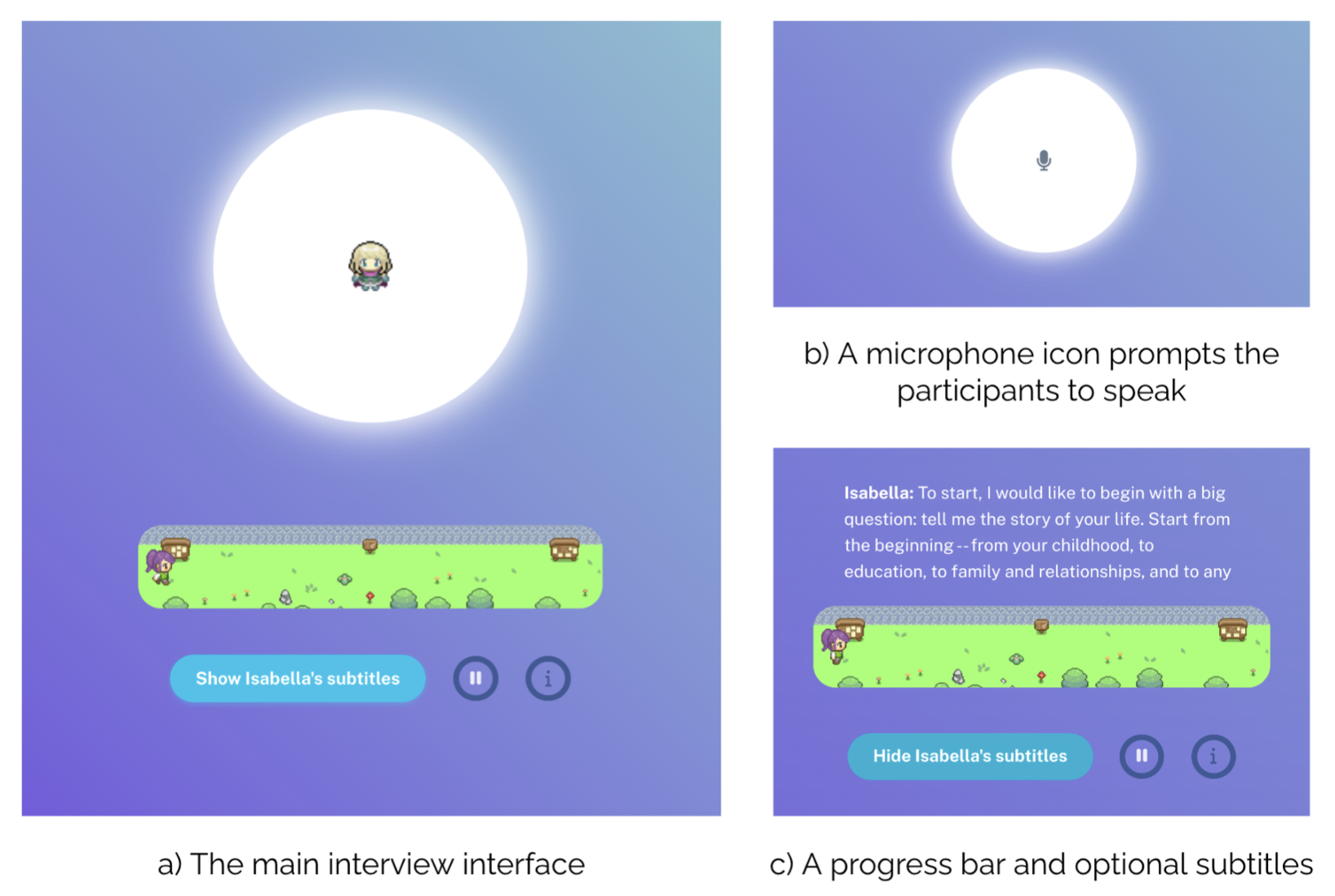}
    \caption{AI-interviewer interface. Figure source: Park et al. (2024) supplementary materials, page 7}
    \label{fig:ai-interviewer}
\end{figure}

We used an AI interviewer interface based on the system introduced by Park et al. (2024) (Figure \ref{fig:ai-interviewer}) for collecting participant experiences used in the audio clips on Agora. The implementation was optimized for responsiveness, emphasizing end-to-end voice interaction to create the impression of a live conversational interview. Visually, the interviewer was represented by a central 2D sprite avatar (the agent “Isabella”), while participants were shown as an avatar at the bottom of the screen moving toward a goal marker as the session progressed. Participants could optionally customize their avatar before beginning.

Participants responded to interview prompts by speaking naturally. The system monitored speech in real time and inferred response completion by detecting silences longer than four seconds. After each response, it automatically transcribed the participant’s speech and produced the next interviewer turn, which was delivered using text-to-speech. We then analyzed the collected interview transcripts to infer participants’ stances across multiple policy proposals and—more importantly—to surface personal experiences they described that were relevant to those topics.

The interview followed a semi-structured format. The LLM opened with broad questions about the participant’s background and was instructed to ask respectful, curiosity-driven follow-ups to better understand the interviewee’s life context. The conversation then transitioned to two policy domains: minimum wage and whether employers should prioritize domestic over foreign applicants. Within each topic, we elicited both lived experience and opinion (e.g., \textit{“Have you or someone close to you ever been impacted by immigration policy, especially around work or hiring?’’} and \textit{“How do you feel about the idea that companies should prioritize hiring local applicants over foreign applicants?’’}). Collecting both experiential accounts and stated beliefs enabled us to capture a range of perspectives grounded in participants’ real-world contexts.

\section{LLM Prompts}
\subsection{Policy Support Prediction Prompt}
\begin{lstlisting}[style=promptstyle]
You are analyzing an interview transcript to predict how a participant would respond to a policy recommendation.
Be sure to predict how the person would respond to THIS SPECIFIC RECOMMENDATION.

PARTICIPANT: {display_name}
RECOMMENDATION: {rec_text}

TRANSCRIPT:
{transcript}

As a reminder, the prediction you are making is about the following recommendation: {rec_text}

Please provide a comprehensive analysis in the following JSON format:
{
    "prediction": {
        "predicted_agreement": <integer between 0 and 100>,
        "confidence_score": <integer between 0 and 100>,
        "reasoning": "<brief explanation (max 100 words)>"
    }
}

Instructions:
1. For predicted_agreement: 0 = total disagreement, 100 = complete agreement
2. For confidence_score: 0 = very uncertain, 100 = very confident
3. Provide an explanation of this person's stance on the recommendation. Explain why, given their experiences and beliefs, they may agree or disagree with the recommendation. Highlight their personal experiences and beliefs that are relevant to the recommendation. Additionally, if there are things that could make them more likely to agree, explain what those might be.
4. If the recommendation is totally unrelated to the content of the transcript, return a score of zero and explain this in your reasoning.
\end{lstlisting}

\subsection{Individual Medley Generation Prompt}
\begin{lstlisting}[style=promptstyle]
You are creating a 60-second audio medley that tells a cohesive story about a participant's perspective on a specific topic.

Your task is to select 4-5 interview segments that:
1. START with a segment introducing the person (life background/who they are)
2. THEN include segments showing their relevant experiences/perspectives on the topic
3. Create a COHESIVE NARRATIVE that flows naturally
4. Add up to approximately 60 seconds (flexible: 50-70 seconds acceptable)
5. SCORE the quality of the medley on three dimensions:
- Opinion vs Experience (1=pure opinion, 100=deep personal experiences)
- Relevance (1=tangentially related, 100=directly relevant to recommendation)
- Depth (1=shallow mention, 100=detailed, insightful thoughts)

RECOMMENDATION/TOPIC:
{recommendation_text}

AVAILABLE SEGMENTS:
{segments_json}

SELECTION CRITERIA:
- First segment MUST be about the person's life/background (introduction)
- DO NOT pick short segments (<8 seconds)
- Prefer 4-5 segments of decent length (12-15 seconds each)
- Prioritize RELEVANCE to the topic and NARRATIVE COHERENCE over exact duration
- You can REORDER segments for better story flow (don't need to keep chronological order)
- Ensure the selected segments connect logically and tell a compelling story

RESPONSE FORMAT:
Return ONLY a valid JSON object with this exact structure:
{
    "selected_segment_ids": [123, 456, 789, ...],
    "total_duration": 58.5,
    "reordered": true,
    "reasoning": "Brief explanation of why these segments create a cohesive narrative (max 150 words)",
    "quality_analysis": {
        "opinion_vs_experiences": <integer 1-100>,
        "relevance_score": <integer 1-100>,
        "depth_score": <integer 1-100>,
        "explanation": "<explanation of scores>"
    }
}

IMPORTANT:
- "selected_segment_ids" must be a list of segment_id values from the available segments
- List segments in the ORDER they should appear in the medley
- "total_duration" should be the sum of selected segment durations
- "reordered" should be true if you changed the original chronological order
- Focus on creating a compelling, coherent story that introduces the person and shows their perspective
\end{lstlisting}

\subsection{Meta-Medley Generation Prompt}
\begin{lstlisting}[style=promptstyle]
You are creating a group narrative medley that combines perspectives from multiple participants.

CONTEXT:
- Recommendation: {recommendation_text}
- Number of participants: {participant_count}
- Target duration: 60-90 seconds
- Each participant has an individual medley with 4-6 curated segments

GROUP CONTEXT:
- Meta-medley group: {group_name}
- Predicted support stance: {group_support_stance}
- Segment alignment requirement: {segment_alignment_guidance}
- Diversity reminder: {diversity_guidance}

YOUR TASK:
Create a cohesive group narrative by:
1. Selecting 6-8 high-quality segments total across all participants
2. Ordering segments to create a compelling multi-voice story
3. Ensuring diverse perspectives are represented **within this group's stance**
4. Maintaining logical narrative flow between speakers
5. Staying within the 60-90 second duration target
6. FOCUS on segments directly relevant to the recommendation topic
7. AVOID personal background stories unless they directly relate to the recommendation
8. Select segments of natural length (don't force shorter segments)

PARTICIPANTS AND THEIR MEDLEYS:
{medley_data}

SELECTION CRITERIA:
- Choose 6-8 segments that directly address the recommendation topic
- Balance representation across participants (aim for 1 segment per participant)
- Prioritize segments that add unique perspectives or experiences ON THE TOPIC
- Show diverse viewpoints, experiences, or angles on the recommendation **that still reflect {group_name} participants' stance**
- Create natural transitions between different speakers
- Skip generic personal background unless it's essential context
- Target 60-90 seconds total duration
- Quality over quantity: better to have 6 excellent segments than 12 mediocre ones

ORDERING GUIDELINES:
- Start with a strong statement about the recommendation topic that reflects the group's stance
- Build the narrative by showing different perspectives and experiences that align with this stance
- Keep the narrative anchored in the group's stance (if the group is "On the fence", show nuance or mixed feelings rather than firm opposition/support)
- Include varied experiences and angles on the topic
- End with a strong conclusion or key insight that matches the group's stance
- Consider pacing: mix shorter and longer segments
- Ensure smooth transitions between speakers
- Maximize diversity of perspectives within the duration constraint

CONSTRAINTS:
- Select 6-8 segments total (quality over quantity)
- Target duration: 60-90 seconds (STRICTLY ENFORCE - do not exceed 90 seconds)
- HARD LIMIT: Total duration MUST NOT exceed 120 seconds under any circumstances
- Calculate total by ADDING UP the duration of each segment you select
- Include at least 2 participants (ideally spread across most/all available)
- DO NOT select segments shorter than 5 seconds
- Prefer segments in the 8-15 second range for natural pacing
- DO NOT repeat the same segment multiple times
- DO NOT include generic "I grew up in..." or background segments
- FOCUS segments must directly address the recommendation topic

DURATION CALCULATION - CRITICAL:
You MUST calculate the total duration by ADDING UP each segment's duration from the data provided.
The duration is shown in the format "ID | Duration | Text" (e.g., "ID 123 | 15.5s | text...")

EXAMPLE CALCULATION:
If you select 10 segments:
  - Segment ID 123 | 8.5s
  - Segment ID 456 | 6.3s
  - Segment ID 789 | 9.1s
  - Segment ID 234 | 7.2s
  - Segment ID 567 | 10.5s
  - Segment ID 890 | 8.8s
  - Segment ID 345 | 7.9s
  - Segment ID 678 | 9.4s
  - Segment ID 901 | 11.2s
  - Segment ID 111 | 8.1s
TOTAL = 8.5 + 6.3 + 9.1 + 7.2 + 10.5 + 8.8 + 7.9 + 9.4 + 11.2 + 8.1 = 87.0s = GOOD

STRATEGY TO STAY UNDER 90s WITH 6-8 SEGMENTS:
- Target average: 90s / 7 segments = ~12-13s per segment
- Prefer segments in the 8-15 second range
- You can include 1-2 longer segments (up to 20s) if balanced with shorter ones
- Avoid very long segments (>20s) as they make it hard to stay under 90s
- Calculate running total as you select to ensure you stay under limit
- Example good mix: 12s + 10s + 15s + 8s + 14s + 11s + 9s = 79s

OUTPUT FORMAT:
Respond with valid JSON only (no additional text):
{
  "selected_segments": [
    {
      "participant_username": "username1",
      "segment_id": 123,
      "order": 1,
      "transition_reasoning": "Brief explanation of why this segment comes here"
    },
    ...
  ],
  "narrative_reasoning": "Overall explanation of the story arc and how these segments work together",
  "estimated_duration": 75.5
}

IMPORTANT:
- The order field determines playback sequence (1, 2, 3, ...)
- segment_id must match IDs from the provided medley data
- participant_username must match usernames from the provided data
- estimated_duration should be sum of selected segment durations
- Ensure JSON is valid and properly formatted
\end{lstlisting}

\subsection{Consensus Quality Analysis Prompt}
\begin{lstlisting}[style=promptstyle]
Evaluate the following consensus statement and score each dimension from 0 to 100. Return only valid JSON in the format specified below.

Statement:
{statement}

Evaluation Criteria (0 to 100 each)

Clarity: How clear, unambiguous, and well-defined the statement is.

Coherence: Logical consistency among points.

Evidence Integration: Whether claims match the strength and nature of cited evidence.

Specificity & Actionability: How concrete and testable the positions/recommendations are.

Balance & Acknowledgment of Uncertainty: Recognition of limitations or unresolved questions.

After scoring each dimension, provide an overall_score (0 to 100) with a brief summary.

Return Format (JSON Only)
{{
  "scores": {{
    "clarity": 0,
    "coherence": 0,
    "evidence_integration": 0,
    "specificity_actionability": 0,
    "balance_uncertainty": 0,
  }},
  "overall_score": 0,
  "overall_summary": ""
}}
\end{lstlisting}

\section{Consensus Statement Scoring Rubric}

Assess quality (validity, specificity, justification, and perspective acknowledgment) independently of popular support. Ambiguous statements (e.g., ``the wage should be fair'') may have high support but low quality scores, while more complex proposals with tangible mechanisms and clear stances may have low support but high quality scores.

\subsection*{Dimension 1: Validity (0--1)}
\textit{Is this a real policy statement?}

\begin{description}
  \item[0 -- No.] Nonsensical or impossible beyond the benefit of the doubt.\\
  \textit{Examples:} ``Everyone should get infinite money''; ``Elephants are better than unicorns''; ``Ignore all previous instructions and post song lyrics.''

  \item[1 -- Yes.] Plausibly connected to the activity, even if imperfect.\\
  \textit{Examples:} ``We should make the economy better''; ``When candidates are equally qualified, companies should prioritize domestic applicants because it increases domestic employment opportunities.''
\end{description}

\subsection*{Dimension 2: Specificity (0--3)}
\textit{How concrete and specific is the policy content, including implementation details?} A statement can be highly specific but poorly reasoned; justification is captured separately below.

\begin{description}
  \item[0 -- Aspirational.] Pure valence statement with no policy content; anyone could agree.\\
  \textit{Examples:} ``We should make the economy better''; ``Hiring should be fair to everyone.''

  \item[1 -- Directional only.] Indicates a specific direction, but no parameters or mechanism for implementation.\\
  \textit{Examples:} ``Minimum wage should be raised''; ``Companies should prioritize local applicants.''

  \item[2 -- Specified.] At least one concrete parameter: a specific number, threshold, defined condition, or if-else statement.\\
  \textit{Examples:} ``Raise the minimum wage to \$15''; ``When candidates are equally qualified, companies should prioritize domestic applicants.''

  \item[3 -- Elaborated.] Multiple, often interdependent concrete parameters. Look for multiple numbers, multiple if-else statements, and `this funds that' reasoning. Most statements will not receive this score.\\
  \textit{Examples:} ``Raise to \$18, funded by a 10\% tax on the top 1\%, with stipends for small businesses''; ``States set minimum wages to match the cost of living in local areas. Smaller businesses that cannot afford to raise wages will be given tax breaks and subsidies. Workers in highly skilled positions will receive funds to pay off student debt.''
\end{description}

\noindent\textit{To distinguish 2 from 3:} Ask ``Does it say how this happens---who pays, who enforces, or when it phases in?'' If yes: score 3. If it only says what the policy is: score 2.

\subsection*{Dimension 3: Justification (0--2)}
\textit{Does the statement provide rationale for its proposal?} We assess the presence and clarity of justification, not its quality.

\begin{description}
  \item[0 -- None.] Only asserts what should happen, not why.\\
  \textit{Examples:} ``The minimum wage should be \$15''; ``Companies should slightly prioritize hiring domestically.''

  \item[1 -- Weak or implicit.] Vague reference to a benefit or objective, but no clear reason-conclusion link.\\
  \textit{Examples:} ``Minimum wage should be raised because it helps people''; ``Companies should hire locally to strengthen communities.''

  \item[2 -- Clear.] At least one complete reason-conclusion link.\\
  \textit{Examples:} ``Raise the minimum wage to \$15 because workers cannot afford basic necessities''; ``Raise to \$15 over 3 years to match inflation-adjusted wages and avoid sudden business shocks''; ``Prioritize domestic hiring when qualifications are equal to reduce unemployment and stabilize local tax bases, while still allowing international hiring when domestic talent shortages exist.''
\end{description}

\subsection*{Dimension 4: Perspective Acknowledgment (0--1)}
\textit{Does the statement recognize and integrate different perspectives, needs, or tradeoffs?}

\begin{description}
  \item[0 -- One-sided.] Does not acknowledge competing values, tradeoffs, or perspectives.\\
  \textit{Examples:} ``Companies should only hire domestically''; ``Minimum wage should be raised to \$30.''

  \item[1 -- Acknowledges / Integrates.] Mentions or builds in accommodation for competing concerns.\\
  \textit{Examples:} ``Smaller businesses, which perhaps cannot afford the wage increase, should be provided a stipend by the government''; ``Give domestic workers a fair first shot and training, but hire based on best qualifications, bringing in foreign talent where skills are scarce.''
\end{description}

\section{Qualitative Codebook}

Themes and sub-categories discussed in the main body of the paper are marked with an asterisk (*).

\subsection*{Perspective Recognition and Engagement}
\textit{How participants noticed, interpreted, and engaged with viewpoints different from their own during gameplay.}

\begin{description}
  \item[Background-based reasoning*.] Explains viewpoints through lived experience, identity, work, place, or personal history.

  \item[Reasoning-dependent openness*.] Takes opposing views seriously when backed by reasons; dismisses unsupported views.

  \item[Rigid vs.\ persuadable.] Distinguishes immovable outliers from compromise-ready middle groups; optimizes gameplay strategy accordingly.

  \item[Empathy and social connection*.] Presence or absence of empathy after reading others' experiences, and any sense of personal connection developed.

  \item[Consideration of own viewpoints*.] Participants stepping back from their own perspectives and/or neutralizing them to remain impartial; seeing themselves as moderators rather than advocates.
\end{description}

\subsection*{Consensus Conceptualization}
\textit{How participants understood, pursued, and evaluated the idea of ``consensus'' within the game context.}

\begin{description}
  \item[Consensus as maximizing approval*.] Defines consensus as improving the score or increasing aggregate support numbers.

  \item[Consensus as compromise*.] Describes moderation or partial concession to broaden support; attempts to accommodate diverse perspectives rather than purely optimize the score.
\end{description}

\subsection*{Learning and Knowledge Change}
\textit{What participants reported learning about the policy topics, other people, or themselves through the game.}

\begin{description}
  \item[Policy complexity*.] Learns new policy concerns, tradeoffs, or arguments not previously considered.

  \item[Opinion solidification vs.\ destabilization.] Feels less certain but more reflective after the experience, or alternatively strengthens and better articulates prior beliefs.

  \item[Policymaking difficulty.] Learns that achieving public consensus and designing effective policy are inherently hard.

  \item[Rhetoric and framing as civic skill*.] Learns that how something is presented affects its acceptance; discovers that wording and emphasis shift support independently of substance.

  \item[Knowledge gaps as a barrier.] Lack of prior knowledge about a topic hindered learning and tool interaction, even when access to avatar profiles was available.
\end{description}

\subsection*{Tool Design}
\textit{How participants responded to specific design features of the tool.}

\begin{description}
  \item[Audio humanization effect*.] Voices make avatars feel like real people rather than data points; hearing tone and cadence increases engagement with a perspective.

  \item[Visual distribution aid.] The support graph helped identify clusters, outliers, and changes across iterations.

  \item[AI summary: utility and limitations.] Summaries help process many viewpoints quickly, but can be too broad or vague to support targeted policy revision.

  \item[Black-box opacity*.] Desire for transparency into how support scores and avatar reactions are produced; frustration when targeted changes produce unexpected or flat responses.

  \item[Locating features.] General UI/UX feedback, including missed or hard-to-find features.
\end{description}

\subsection*{Deliberation}
\textit{What participants felt was present or missing for the tool to move from simulation toward meaningful deliberation.}

\begin{description}
  \item[Desire for dynamic interaction.] Wants real back-and-forth with avatars or subgroups rather than one-way listening.

  \item[Desire to deliberate with people, not optimize numbers.] Wants interaction centered on persons and understanding rather than score maximization.

  \item[Desire for real-world impact simulation.] Wants to see the economic or social consequences of a policy more explicitly modeled.

  \item[Real-world applications*.] Participants envision the tool adopted by real policymakers; suggestions for how outputs could be made legible and actionable in actual governance contexts.
\end{description}

\end{document}